\documentclass[journal]{IEEEtran}

\ifCLASSINFOpdf
\else
\fi
\usepackage{mathtools}
\usepackage{cite}
\usepackage[hidelinks]{hyperref}
\usepackage{graphicx, caption, subcaption}
\usepackage{amsmath,amssymb,amsfonts}
\usepackage{algorithmic}
\usepackage{graphicx}
\usepackage{xcolor}
\usepackage[version=4]{mhchem}
\usepackage{textcomp}
\usepackage{siunitx}
\usepackage[utf8]{inputenc}
\usepackage{float}

\usepackage{fancyhdr,lipsum} 
\makeatletter
\def\ps@IEEEtitlepagestyle{ 
	\def\@oddfoot{\mycopyrightnotice}
}
\def\mycopyrightnotice{
	{\scriptsize © 2023 IEEE. Personal use is permitted, but republication/redistribution requires IEEE permission.See https://www.ieee.org/publications/rights/index.html for more information.\hfill}
}
\begin{document}

\title{$H$-$\phi$ Formulation in Sparselizard Combined With Domain Decomposition Methods for Modeling Superconducting Tapes, Stacks, and Twisted Wires}
%
%
%

\author{N. Riva, A. Halbach, M. Lyly, C. Messe, J. Ruuskanen, and V. Lahtinen \thanks{This research is supported by Quanscient, Type One Energy, PSFC MIT, CFS and LBNL. M. Lyly acknowledges support from the Academy of Finland project 324887.}
\thanks{A. Halbach, M. Lyly, and V. Lahtinen are with Quanscient Oy (Tampere, Finland) (e-mail:alexandre.halbach@quanscient.com, 
 mika.lyly@quanscient.com, valtteri.lahtinen@quanscient.com)}
\thanks{M. Lyly and J. Ruuskanen are with Tampere University (Tampere, Finland) (e-mail: mika.lyly@tuni.fi, janne.ruuskanen@tuni.fi)}
\thanks{N. Riva is with MIT Plasma Science and Fusion Center (MA, USA) (e-mail:nicoriva@mit.edu)}
\thanks{C. Messe and J. Ruuskanen are with Lawrence Berkeley National Laboratory (CA, USA) (e-mail:cmesse@lbl.gov)}}

\markboth{$H$-$\phi$ Formulation in Sparselizard and Domain Decomposition Methods for Modeling Tapes, Stacks, and Twisted Wires}{}

\maketitle

\begin{abstract}
The growing interest in the modeling of superconductors has led to the development of effective numerical methods and software. One of the most utilized approaches for magnetoquasistatic simulations in applied superconductivity is the $H$ formulation. However, due to the large number of degrees of freedom (DOFs) present when modeling large and complex systems (e.g. large coils for fusion applications, electrical machines, and medical applications) using the standard $H$ formulation on a desktop machine becomes infeasible. The $H$ formulation solves the Faraday's law formulated in terms of the magnetic field intensity $\mathbf{H}$ using edge elements in the whole modeling domain. For this reason, a very high resistivity is assumed for the non-conducting domains, leading to an ill-conditioned system matrix and therefore long computation times. In contrast, the $H$-$\phi$ formulation uses the $H$-formulation in the conducting region, and the $\phi$ formulation (magnetic scalar potential) in the surrounding non-conducting domains, drastically reducing DOFs and computation time. In this work, we use the $H$-$\phi$ formulation in 2D for the magnetothermal (AC losses and quench) analysis of stacks of REBCO tapes. The same approach is extended to a 3D case for the AC loss analysis of a twisted superconducting wire. All the results obtained by simulations in \texttt{Sparselizard} are compared with results obtained with COMSOL. Our custom tool allows us to distribute the simulations over hundreds of CPUs using domain decomposition methods, considerably reducing the simulation times without compromising accuracy. 
\end{abstract}


\section{Introduction}
\label{sec:introduction}
\IEEEPARstart{W}{hen} modeling superconducting materials, the electrical resistivity is generally modeled using the power law constitutive relationship \cite{Bruzzone2004}, which may include a complex critical current density dependence~\cite{Robert2019}. The highly nonlinear properties and strong anisotropic field dependence of the critical current density could lead to a very large computation time. Moreover, the high aspect ratio of the superconducting tapes (especially in the case of High-Temperature Superconductors (HTS)) leads to a large number of elements and degrees of freedom (DOFs). 
A widely used method is the $H$ formulation~\cite{Brambilla2007}. However, due to the large scale of systems such as electrical machines~\cite{Benkel2020} and fusion devices~\cite{Sorbom2015}, the computational limits of are rapidly reached with the $H$ formulation~\cite{9097858,Shen_2020}. Moreover, the use of the $H$ formulation in nonconducting domains leads to unnecessary large number of DOFs due to the vectorial nature of the magnetic field intensity $\mathbf{H}$ and to numerical instabilities due to the imposed high resistivity to avoid eddy currents in such domains, leading to an ill-conditioned matrix. The development of approaches more efficient than the $H$ formulation to be implemented in commercial and in-house software is of paramount importance to improve the computational efficiency of the models.
Recently, several works have led to drastic improvements in computational efficiency using the $A$-$H$~\cite{bortot2020coupled,Dular.2019, Dular2021}, the $T$-$A$ (similar to the $A$-$H$)~\cite{zhang2016efficient,liang2017finite,grilli2020electromagnetic}, and the $H$-$\phi$ formulations~\cite{lahtinen2015, stenvall2014, Arsenault.2020, arsenault2021comsol}.

This paper aims at addressing the current challenges of 3D modeling 2G HTS using the $H$-$\phi$ formulation combined with domain decomposition methods (DDM)~\cite{schwarz1869ueber,lions1988schwarz}, enabling massive parallel computation and drastically reduced simulation time. The presented case studies are chosen to represent fusion-energy inspired industrially relevant cases in AC loss and quench modeling.

In section~II, we briefly describe the formulation and its implementation in \texttt{Sparselizard} and we describe the utilized custom DDM tool. In section~III, we present validating results using simple 2D models, and in section~IV, we move on to more complex 3D models, demonstrating the virtues of our DDM-based tool. Finally, in section~V, we draw conclusions.      

\section{H-$\phi$ Formulation and Implementation}
\label{sec:theoryMaterial}

\subsection{Formulation}
The $H$-$\phi$ formulation where current constraints are imposed using cohomology cuts is well-known in computational electromagnetics \cite{Kettunen1998a, Kettunen1998b, Henrotte2003, Pellikka2013} and the mathematical main ideas in an electromagnetic context can be traced back to Kotiuga's early works on making cuts for scalar potentials \cite{Kotiuga1987}. Eventually, it was brought to the context of superconductor AC loss simulations by Lahtinen, Stenvall {\it et al.} \cite{lahtinen2015, stenvall2014}.

The finite element formulation is obtained by developing the weak form of Faraday’s law of induction and the Gauss law for the magnetic field. Ohm’s law is used as transport law. The magnetic field strength $\mathbf{H}$ is discretized in the conducting regions of the computational domain with Nédélec elements, or Whitney 1-forms, where the degrees of freedom are associated with edges in the first-order interpolation, thus fulfilling Ampere’s circuital law. In regions that are both non-conducting and non-ferromagnetic, Lagrangian elements are used. Their nodal degrees of freedom represent the magnetic scalar potential $\mathrm{\phi}$. 

By exploiting the fact that the curl of the gradient of a scalar field is zero, the term containing the electric resistivity vanishes, thus circumventing the ill-conditioning of the system matrix that would occur if one modeled the air using $\mathbf{H}$ directly and imposed a very high electric resistivity. The benefit of using the magnetic scalar potential $(\mathrm{\phi})$ rather that the vector potential A is that only one degree of freedom per node is required. This way, the number of unknowns is drastically reduced. The net current constraints are imposed using cohomology cuts, which ensure that circulations of the magnetic field around conducting domains are equal to the desired net currents \cite{Pellikka2013}. Note that this would be impossible using only a gradient of a scalar field, since a circulation of a gradient field over a closed loop is always zero. Hence, we express $\mathbf{H}$ in the nonconducting regions as $\mathbf{H} = -\mathrm{grad} (\phi) + C$, where $C$ is a field associated with the representative of the first cohomology group of the nondconducting region \cite{lahtinen2015}. The external field can be applied as a non-zero Neumann condition on the boundary. A recent overview of the $H$-$\phi$ formulation in applied superconductivity and its numerical stability is given by Dular {\it et.~al.}~\cite{Dular.2019, Dular2021}. Moreover, recent detailed explanations on how the domain interfaces between the conducting and non-conducting regions are formulated, as well how current-boundary conditions are applied are discussed by Arsenault {\it et al.}~\cite{Arsenault.2020, arsenault2021comsol}.
\subsection{Domain Decomposition Method}
The domain decomposition method (DDM) in the FEM context refers to the partitioning of the computational mesh into similarly sized pieces that can each be processed on different computing instances, allowing to distribute the computational burden. The DDM method is used to improve the speed of numerical simulations in solid mechanics, electromagnetism, flow in porous media, etc., on parallel machines from tens to hundreds of thousands of cores. This is well suited to take advantage of supercomputer or cloud architectures. In this work, the optimized Schwarz algorithm is used in the DDM framework: this algorithm solves iteratively the problems defined on the smaller mesh pieces and exchanges boundary data between domains at each iteration to reach convergence. Compared to the pioneering work of H. A. Schwarz \cite{schwarz1869ueber} the method used in this work accelerates convergence using generalized minimal residual method (GMRES) and optimized boundary data \cite{lions1988schwarz}. 
\subsection{Implementation}
The weak formulation required for $H$-$\phi$ is implemented in the open source FEM library \texttt{Sparselizard} available at \url{www.sparselizard.org}. The cohomology cuts are obtained from the mesh generator GMSH \cite{geuzaine2009gmsh}. \texttt{Sparselizard} is designed to solve general weak formulations of PDEs and is therefore suited for the multiphysics simulations required in applied superconductivity. Detailed examples of $H$-$\phi$ formulated AC loss problems can be found online. The resolution is accelerated using domain decomposition on an Amazon Web Services (AWS) cloud infrastructure suited for multiphysics DDM, provided by Quanscient (\url{www.quanscient.com}) under the \texttt{Quanscient.allsolve} software.

\section{Validations and Applications in 2D}
The numerical study is conducted on three different problems, comparing the simulations obtained in \texttt{Sparselizard} with the results obtained with the commercial software COMSOL~\cite{COMSOL}. The 2D simulations used the H-$\mathrm{\phi}$ formulation~\cite{Arsenault.2020,Arsenault.2021}. The first problem consists of an elliptic superconducting tape carrying different sinusoidal transport currents; the second problem consists of a superconducting homogenized stack of tapes carrying a sinusoidal transport current at $0.8\, I_{\rm{c}}$ oscillating at two different frequencies; the third problem is the quench of the cross-section of a VIPER cable~\cite{Hartwig2020}. 
The non-linear $E$–$J$ power-law~\cite{Rhyner1993,Liang2015} characteristic of the HTS materials is modeled as:
\begin{equation}
    \rho_{\rm{PLW}}=\frac{E_{\rm{c}}}{J_{\rm{c}}(B_{\parallel},B_{\perp})}\Bigg(\frac{|J|}{J_{\rm{c}}(B_{\parallel},B_{\perp})}\Bigg)^{n-1},
    \label{eq:ANisotropicKIM}
\end{equation}
where $E_{\rm{c}}=\SI{1}{\micro\volt\per\centi\meter}$ is the electric field criterion, $n$ is the power-law exponent and $J_{\rm{c}}(B,\theta)$ is the anisotropic critical current density model~\cite{thakur2011frequency} used to take into account the dependence of the critical current density on the magnetic field and its orientation. The critical current density $J_{\rm{c}}$ in \eqref{eq:ANisotropicKIM} follows a modified version of the Kim model~\cite{thakur2011frequency} and reads as follows:
\begin{equation}
    J_{\rm{c}}(B_{\parallel},B_{\perp})=\dfrac{J_{\rm{c,0}}}{\Bigg( 1+ \dfrac{\sqrt{k|B_{\parallel}|^2+|B_{\perp}|^2}}{B_{\mathrm{c}}} \Bigg)^b},
\end{equation}
where $B_{\mathrm{c}}$, $k$ and $b$ are constants, and $B_{\parallel}$ and $B_{\perp}$ are, respectively, the parallel and perpendicular components of the magnetic flux density with respect to the tape's flat surface.\\
The AC losses per length are calculated as:
\begin{equation} 
P=\int _{\Omega }\mathbf {E}\cdot \mathbf {J}\ \text {d}\Omega,
\label{eq:LossesInst}
\end{equation}
where $\mathbf {E}$ is the electric field, $\mathbf {J}$ the current density, and $\Omega$ is the superconducting domain. 
\subsection{AC Losses in a Single HTS Tape}
\begin{figure}[tb]
  \includegraphics[width=0.52\textwidth]{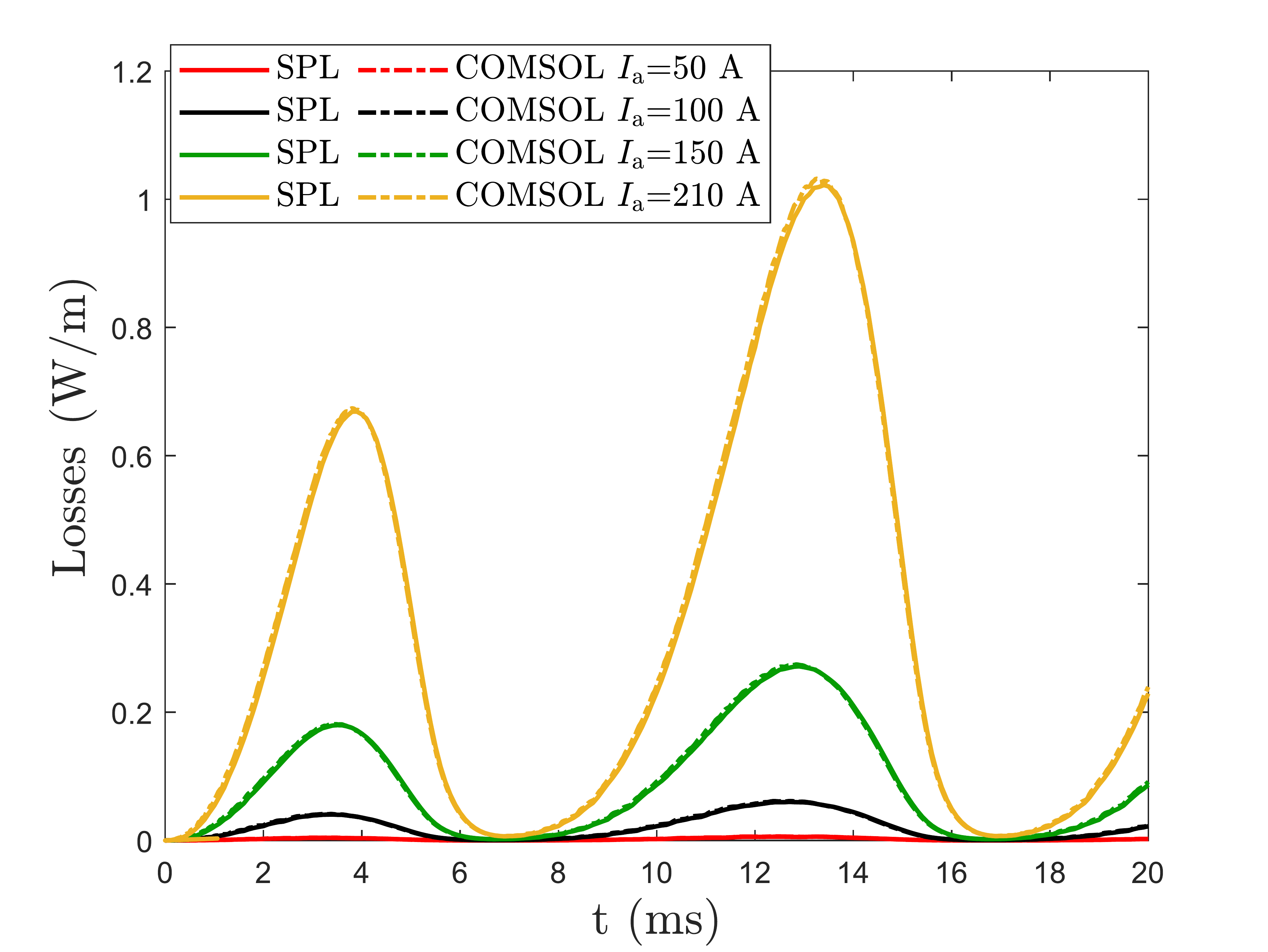}
  \caption{\label{fig:SingleTape} \scriptsize{Instantaneous AC losses in the tape for different transport currents.}}
\end{figure}
For the first model we simulated an infinitely long tape with elliptical cross section carrying different sinusoidal transport currents. The elliptical wire has a cross-section defined by the semi-axes $a=\SI{2}{\milli\meter}$ and $b=\SI{0.08}{\milli\meter}$~\cite{zermeno2013calculation,SuperPower}. The critical current is $I_{\rm{c,0}}=\SI{251}{\ampere}$, with an $n=30$, and the parameters of the anisotropic critical current model are $k=0.25$, $b=0.6$ and $B_{\rm{c}}=\SI{35}{\milli\tesla}$. 
Figure~\ref{fig:SingleTape} shows a comparison of the instantaneous AC losses in the tape for different transport currents calculated with \texttt{Sparselizard} (continuous line) and COMSOL (dashed line). The agreement between all the simulations is excellent. 
\subsection{AC Losses in a HTS Stack}
\begin{figure}[tb]
  \includegraphics[width=0.52\textwidth]{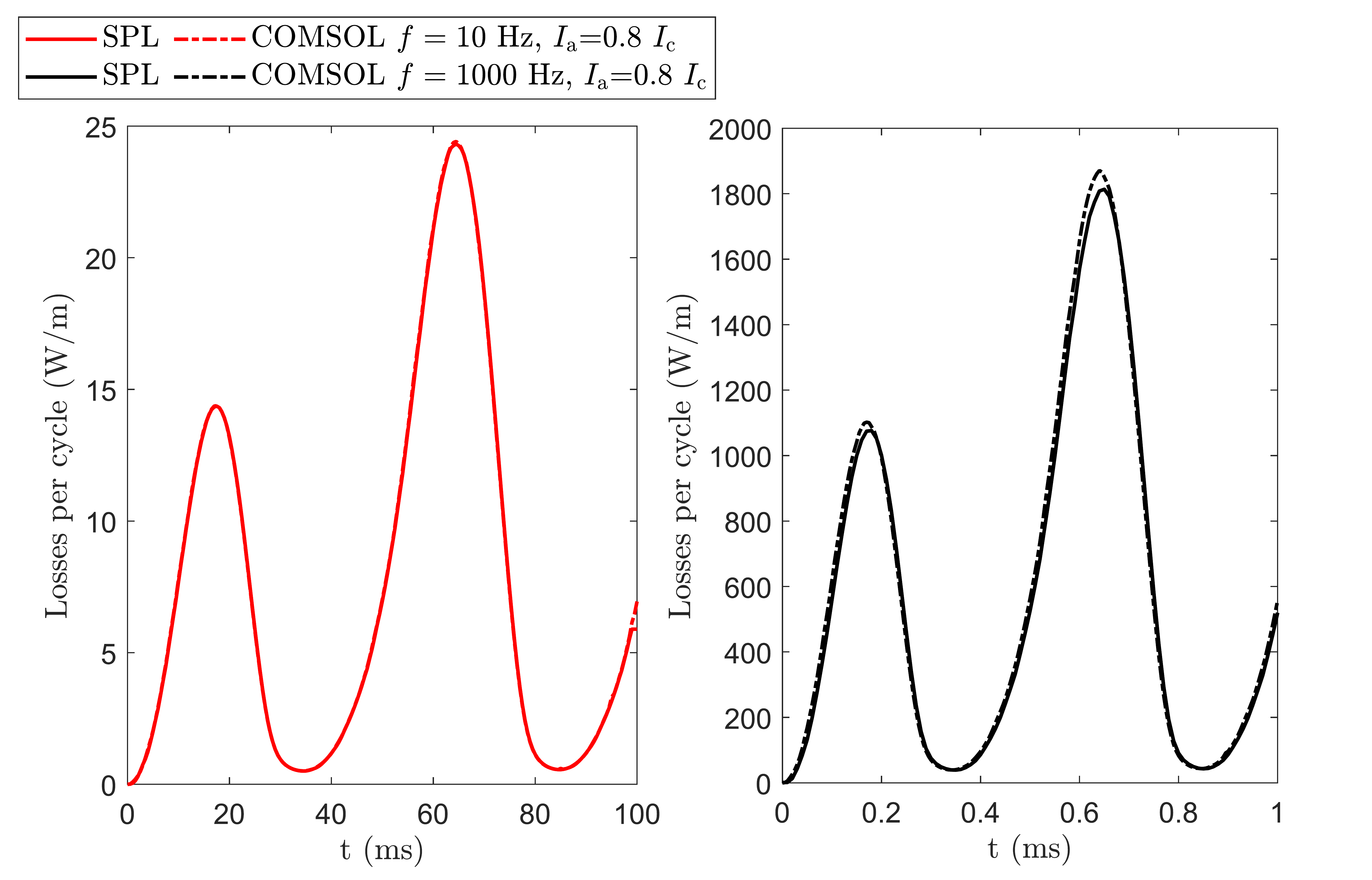}
  \caption{\label{fig:StackTapes} \scriptsize{Instantaneous AC losses in the stack of tapes for $I_{\mathrm{op}}=0.8\, I_{\rm{c}}$ at (left) $f=\SI{10}{\hertz}$ and (b) $f=\SI{1000}{\hertz}$. 
  }}
\end{figure}
The second 2D model consists of infinitely long homogenized anisotropic superconducting bulk \SI{4}{\milli\meter} wide and \SI{4.65}{\milli\meter} high, constituting a bulk of 100 tapes. The $n$-value was $n=30$, and the $J_{\rm{c}}(B_{\parallel},B_{\perp})$ had the same parameters as the single tape, except for $J_{\rm{c0}}$ which was reduced by a factor of 46.5 according to the superconducting bulk homogenization technique~\cite{zermeno2013calculation}. Two case scenario were run: a) $I_{\mathrm{op}}=0.8\, I_{\rm{c}}$ at $f=\SI{10}{\hertz}$ and b) $I_{\mathrm{op}}=0.8\, I_{\rm{c}}$ at $f=\SI{1000}{\hertz}$. The inset of Figure~\ref{fig:StackTapes} represents the geometry and of the normalized critical current density for the selected case of $f=\SI{1000}{\hertz}$ at $t=\SI{5}{\milli\second}$, and a comparison of the instantaneous AC losses in the tape for different transport currents, calculated with \texttt{Sparselizard} (continuous line) and COMSOL (dashed line) using equation~\eqref{eq:LossesInst}. The agreement between the simulations is excellent.
\subsection{Quench in a HTS Stack}
\begin{figure}[tb]
\centering
  \includegraphics[width=0.45\textwidth]{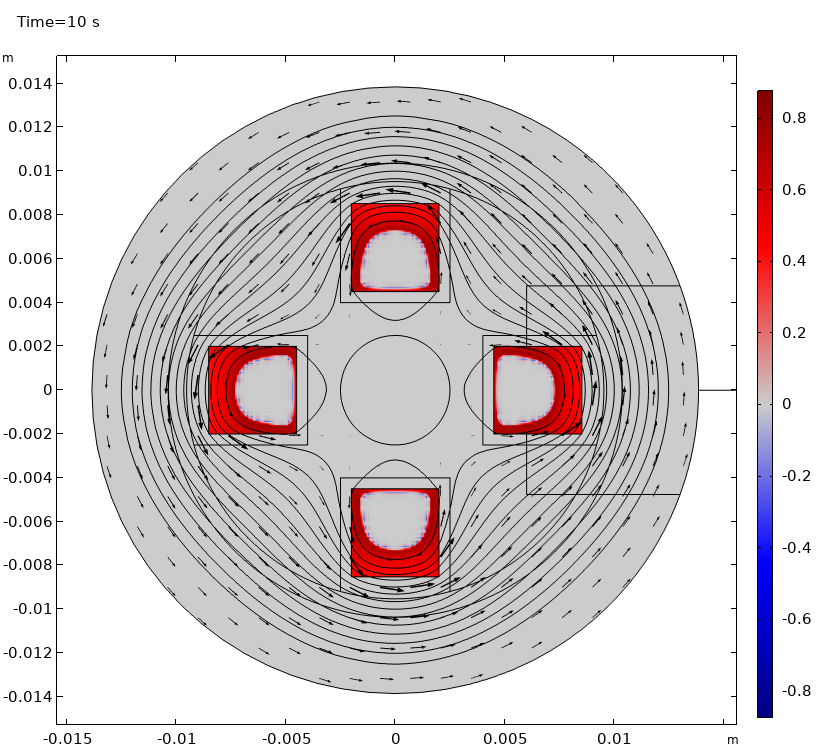}
  \caption{\label{fig:JJcoViper} \scriptsize{Normalized current density normalized for the selected case of $J_{\rm{c}}=J_{\rm{c}}(T,B)$ at $t=\SI{10}{\second}$.}}
\end{figure}
\begin{figure}[tb]
\centering
  \includegraphics[width=0.45\textwidth]{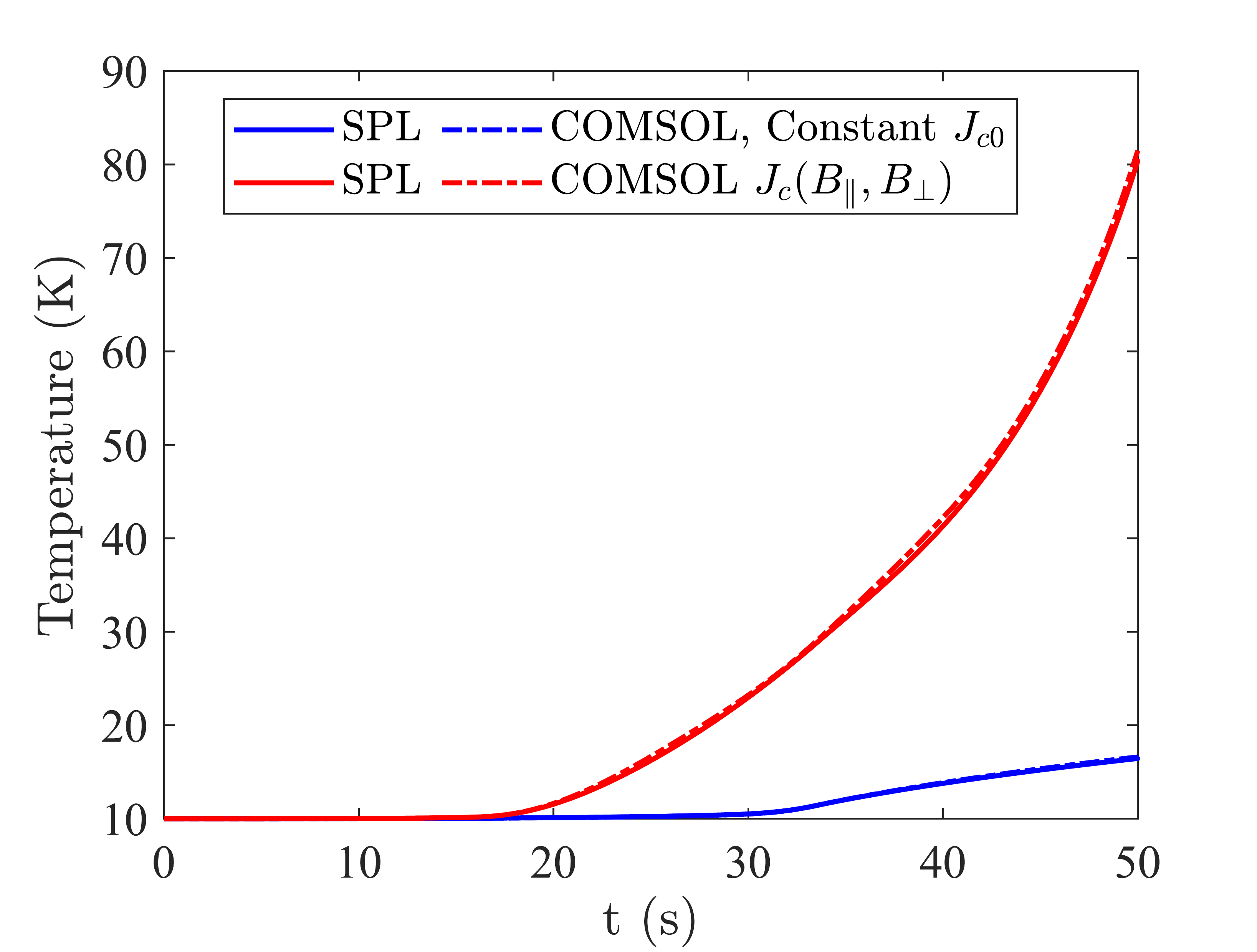}
  \caption{\label{fig:TempQuench2D} \scriptsize{Comparison of the average temperature calculated over the cross section and simulated with \texttt{Sparselizard} (continuous line) and COMSOL (dashed line), for the two scenarios.}}
\end{figure}
The third 2D model consists of a quench simulation of the VIPER cable~\cite{Hartwig2020} where each slot was equipped with a superconducting bulk (each one representing 100 tapes) \SI{4}{\milli\meter} wide and \SI{4.02}{\milli\meter} high, for a total of 4 bulks (400 tapes). Two cases were run; the first one with a critical current density $J_{\rm{c}}(T)$ that decreases linearly with the temperature but does not depend by the applied magnetic field; the second one with a critical current density $J_{\rm{c}}(T,B)$ that depends on both temperature and applied magnetic field~\cite{thakur2011frequency}. The self-field critical current density was $J_{\rm{c0}}=\SI{3.6e10}{\ampere\per\meter\squared}$ and was reduced of a factor 100 (homogenized bulk). This results into a critical current per bulk of $I_{\rm{cb0}}=\SI{5.78}{\kilo\ampere}$ and a total current of the cable of $I_{\rm{cc0}}=\SI{23.1}{\kilo\ampere}$. A current ramp was driven overcurrent at $1.2\,I_{\rm{cc0}}=\SI{27.7}{\kilo\ampere}$, resulting into a uniform quench scenario. A screenshot of the geometry and of the normalized current density for the selected case of $J_{\rm{c}}=J_{\rm{c}}(T,B)$ at $t=\SI{10}{\second}$ is represented in Figure~\ref{fig:JJcoViper}. Figure~\ref{fig:TempQuench2D} shows a comparison of the average temperature calculated over the cross section and simulated with \texttt{Sparselizard} (continuous line) and COMSOL (dashed line), for the two scenarios. The agreement between all the simulations is excellent.

\section{Validations and Applications in 3D Using DDM}

We benchmark our DDM approach in \texttt{Sparselizard} against COMSOL in a 3D AC loss simulation. In \texttt{Sparselizard}, the $H$-$\phi$ formulation is utilized. In COMSOL the analysis is carried out using $H$ formulation~\cite{Brambilla2007}, due to the complexity of implementing a general 3D $H$-$\phi$ formulation using COMSOL.

\subsection{Twisted Superconducting Filaments}
\begin{figure}[tb]
\centering
  \includegraphics[width=0.5\textwidth]{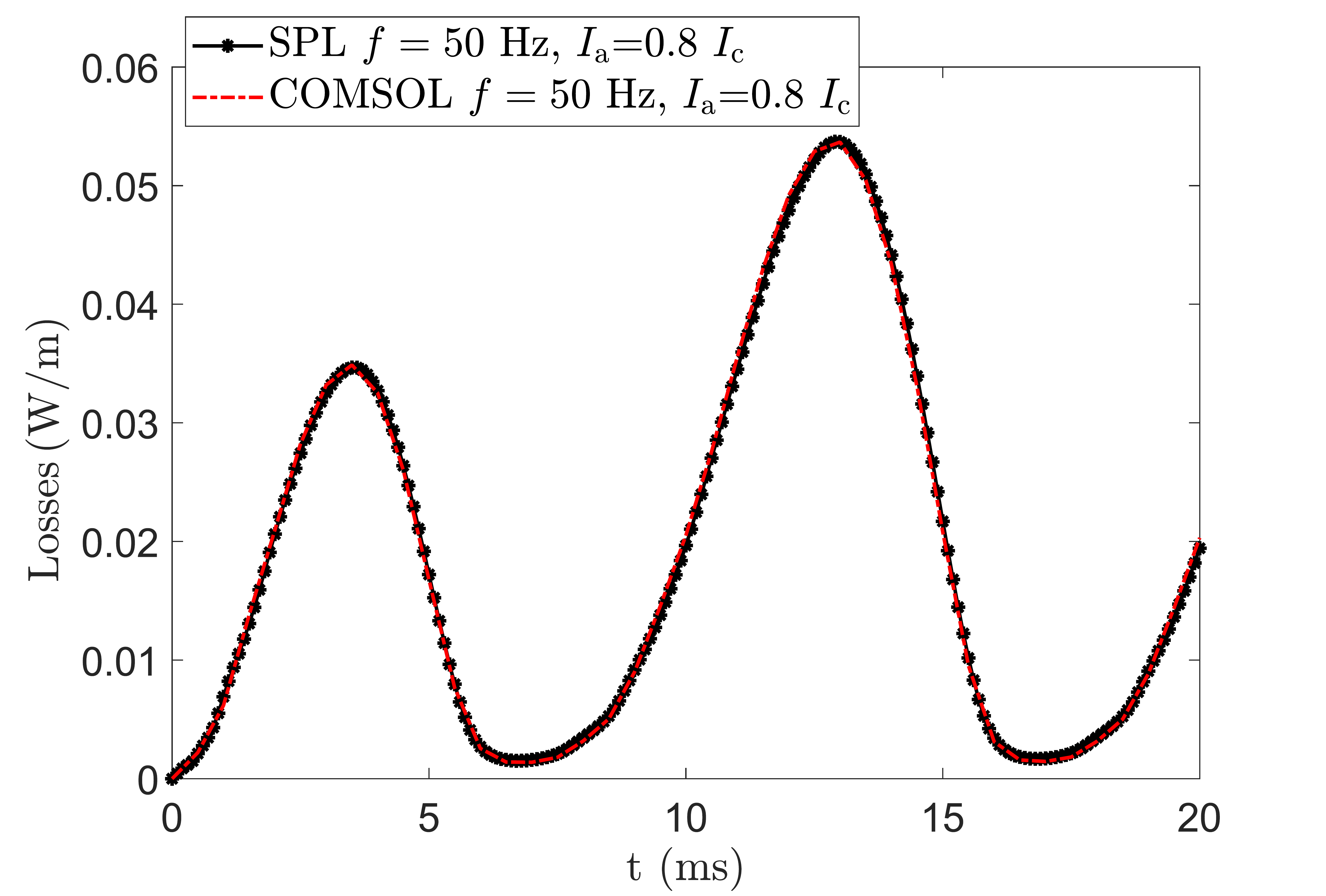}
  \caption{\label{fig:AC_losses_twistedwires} \scriptsize{Instantaneous losses computed on the entire assembly (NbTi wires + copper), with \texttt{Sparselizard} (continuous line + markers) and COMSOL (dashed line).}}
\end{figure}
\begin{figure}[tb]
\centering
  \includegraphics[width=0.425\textwidth]{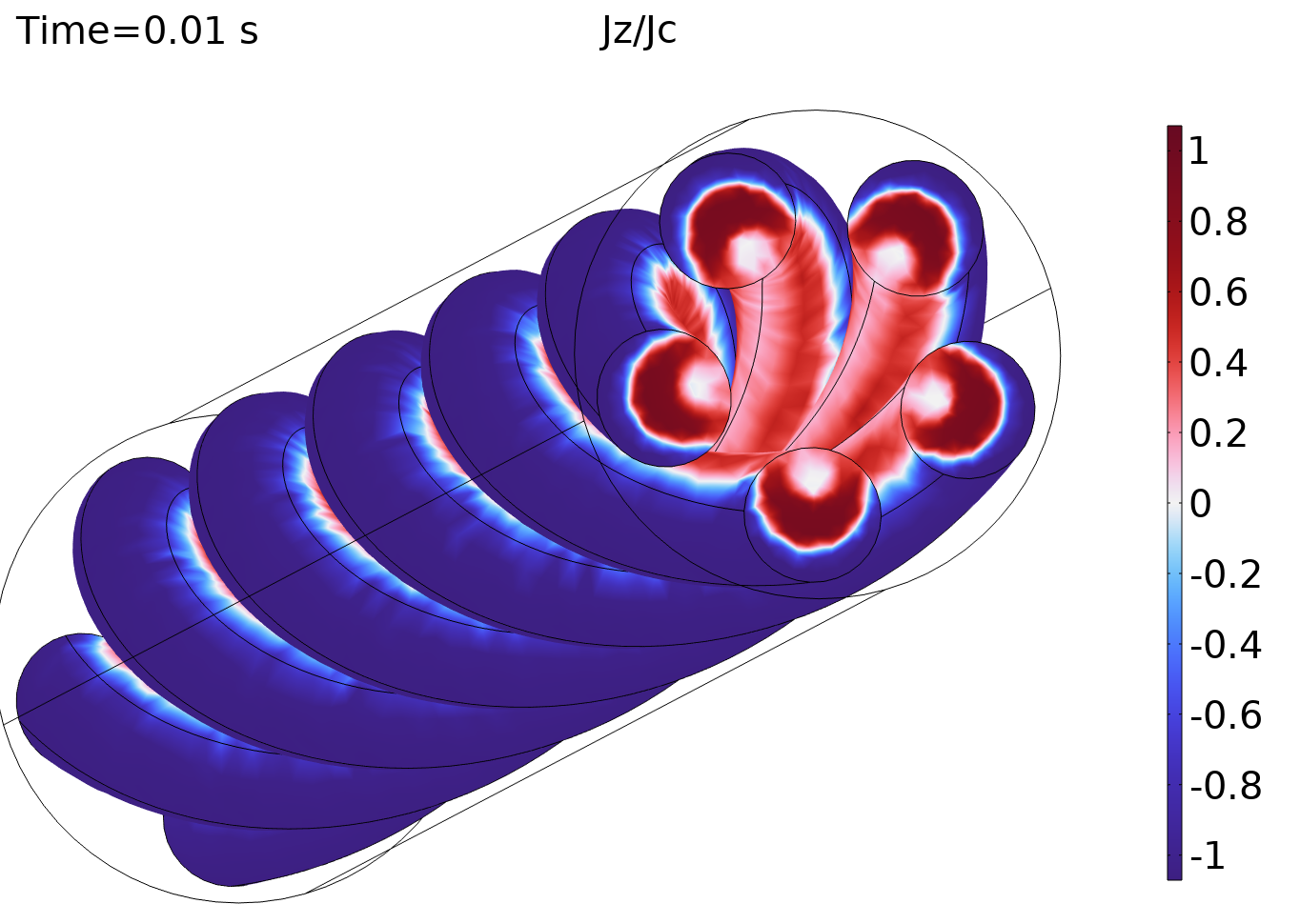}
  \caption{\label{fig:05Jjc_twistedwires} \scriptsize{Normalized current density at $t=\SI{10}{\milli\second}$.}}
\end{figure}
\begin{figure}[tb]
\centering
  \includegraphics[width=0.41\textwidth]{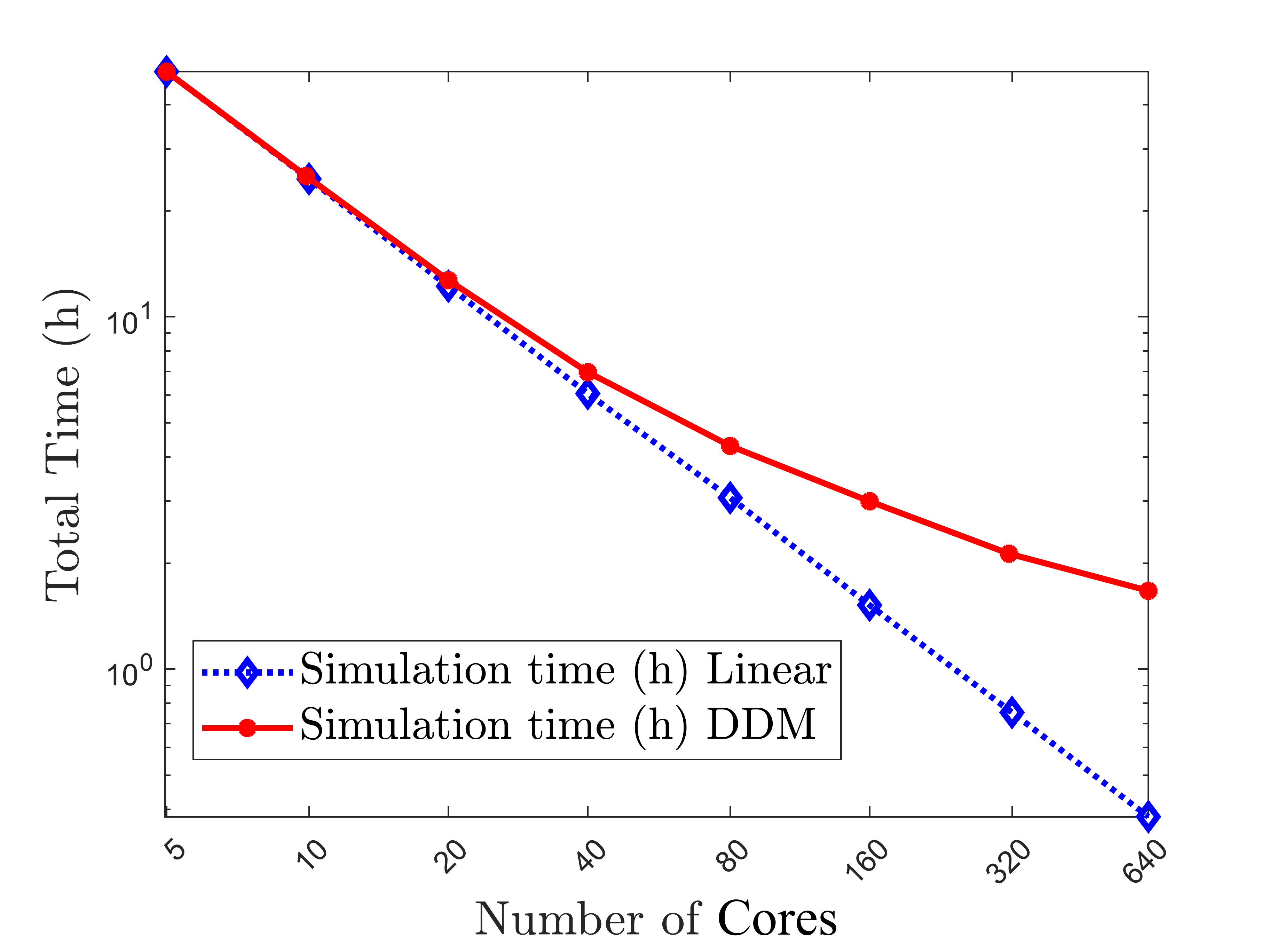}
  \caption{\label{fig:Perf} \scriptsize{Scaling of the computation time of the \texttt{Sparselizard} model on \texttt{Quanscient.allsolve} with the number of single-core CPUs.}}
\end{figure}
The numerical model consists of the 3D simulation of AC losses in a twisted superconducting wire. The wire consists of five homogenized filament bundles of Nb-Ti. Every single filament has a diameter of \SI{350}{\micro\meter} and it is embedded in a copper round core of diameter \SI{535}{\micro\meter}. The power-law constitutive relationship was used with $n=30$ and with a constant critical current of \SI{2.9e8}{\ampere\per\meter\squared}. A full-cycle sinusoidal transport current of amplitude $0.8\,J_\mathrm{c}$ oscillating at $50$~Hz was imposed. Figure~\ref{fig:05Jjc_twistedwires} shows a comparison of the instantaneous AC losses computed on the entire assembly (NbTi wires + copper), with \texttt{Sparselizard} (continuous with markers line) and COMSOL (dashed line) using equation~\eqref{eq:LossesInst}. A screenshot of the geometry and a color map of the normalized current density at $t=\SI{10}{\milli\second}$ is represented in Figure~\ref{fig:AC_losses_twistedwires}. The model run with COMSOL entails 5 MDoFs and it was solved in 7 days, 23 hours, and 40 minutes on a HPC6A server with 96 Cores and 384 GB RAM. In comparison, the $H$-$\phi$-formulation-based model implemented in \texttt{Sparselizard} consists of 1.4 MDoFs and it was solved in 1.7~h using DDM on 640 cores on \texttt{Quanscient.allsolve}. With 80 cores the computation time was 7.2 hours and with 160 cores 4.8 hours. The agreement between the simulation results is excellent. The scaling of the computation time of the \texttt{Sparselizard}-based model on \texttt{Quanscient.allsolve} with the number of CPU cores is shown in Fig.~\ref{fig:Perf}. The number of cores used had no effect on the simulation results.

\section{Conclusions}
Accurate simulation of large-scale AC loss and quench models can take days or even weeks. Therefore, reducing the number of DoFs with clever formulations is pivotal. 
Moreover, parallelizing the analysis further reduces the computation time. In this paper, we compared the $H$-$\phi$-formulated AC loss simulations implemented in \texttt{Sparselizard} with corresponding $H$-$\phi$ and $H$-formulated simulations in COMSOL. In addition, we coupled the $H$-$\phi$ formulation to a thermal problem, simulating a 2D quench in a VIPER cable consisting of HTS stacks. The results show excellent agreement between the different implementations. The discrepancy between simulations is generally lower than 3.5\% (this value was calculated in Fig. 2 for \SI{1000}{\hertz} at \SI{0.64}{\milli\second}). Moreover, we utilized our custom DDM tool to parallelize \texttt{Sparselizard}-based simulations using \texttt{Quanscient.allsolve}. We were able to bring the computation times from more than a week (COMSOL) to just 1.7 hours (\texttt{Sparselizard}) with no loss of accuracy. These results indicate that \texttt{Sparselizard} library combined with the optimized Schwarz DDM is a very efficient and suitable tool for simulations in applied superconductivity.



\bibliographystyle{IEEEtran}
\bibliography{Main}

\end{document}